\documentclass[11pt]{pazh}

\usepackage{psfig}

\begin{document}
        
\setcounter{page}{222}

\sloppypar

\title{\bf HUDF\,1619 -- a candidate polar-ring galaxy in the
Hubble Ultra Deep Field}

\author{V.P. Reshetnikov\inst{1}, R.-J. Dettmar\inst{2}} 

\institute{Astronomical Institute of St.Petersburg State University,
Universitetskii pr. 28, Petrodvoretz, 198504 Russia
\and
Astronomisches Institut der Ruhr-Universit\"at Bochum, Universit\"atsstr. 
150/NA7, 44780 Bochum, Germany}

\authorrunning{Reshetnikov, Dettmar}
\titlerunning{HUDF\,1619}

\abstract{A good candidate for a polar-ring galaxy has 
been detected in the Hubble Ultra Deep Field (HUDF). The galaxy 
HUDF\,1619 ($V\sim25^m$, $z\sim1$) is the most distant object of this type
known to date. A large-scale structure crosses the highly warped disk of 
the main galaxy seen almost edge-on at an angle of about 70$^{\rm o}$. 
The luminosity of this structure (the possible polar ring) reaches  1/3
of the luminosity of the central galaxy. A strong absorption lane is seen 
in the region where this structure is projected onto the disk of the 
central object. There are two galaxies of comparable luminosity adjacent to
HUDF\,1619 (in projection). One of them may be the donor galaxy in the 
interaction which gave rise to the ring structure.
}
\titlerunning{HUDF\,1619}
\maketitle

\section{Introduction}

   Polar-ring galaxies (PRGs) belong to a very rare
and, in many respects, interesting class of objects
(Whitmore et al. 1990). The unusual structure of such
galaxies is commonly attributed to external accretion
and interaction between galaxies. Therefore, if these
processes were more frequent in the Universe in the
past, one may expect the fraction of PRGs to increase
with redshift. Only two candidates for distant objects
of this type detected in the Northern Hubble Deep
Field are known to date (Reshetnikov 1997).

   In this paper, we describe a galaxy found in the
Hubble Ultra Deep Field. Judging by the set of its
morphological and photometric features, it is a good
candidate for a distant ($z\sim1$) PRG. The detection of
this object is consistent with the assumption that the
space density of the galaxies of this type increases
with $z$.

\section{The galaxy HUDF\,1619 in the Hubble Ultra Deep Field}

\subsection{The Hubble Ultra Deep Field}

   The Hubble Ultra Deep Field (its standard abbreviation is 
HUDF) is the deepest optical image of a sky
region obtained to date (Beckwith et al. 2006). The
HUDF (its area is about 11 sq. arcmin.) was observed with the
Hubble Space Telescope using the Advanced Camera
for Surveys (ACS) in the F435W ($B_{435}$), F606W
($V_{606}$), F775W ($i_{775}$), and F850LP ($z_{850}$) broadband
filters (the three digits in the filter names indicate
the central wavelength in nanometers). Several hundred HUDF 
frames were taken from September 2003 through January 2004; 
the total exposure time was about 40$^h$ in the $B_{435}$ and 
$V_{606}$ filters and almost 100$^h$ in the $i_{775}$ and z850 filters.
                                           
     The resulting HUDF images at 0.$''$03 steps are
publicly accessible at the site of the Space Telescope
Science Institute (STScI). Objects as faint as $\sim30^m$
are identified in the field; the actual angular resolution
is $\sim 0.''1$.

\subsection{The galaxy HUDF\,1619}

   The optical morphology of HUDF 1619 (the number in the name 
is the galaxy number in the catalog by
Beckwith et al. (2006) for the $i_{775}$ band) is typical of
PRGs (Figs. 1 and 2): a large-scale structure seen
almost edge-on crosses the stellar disk also seen almost edge-on 
at an angle of about 70$^{\rm o}$. An absorption
lane is clearly seen in the region where the inclined
structure crosses the disk of the central galaxy. Both,
the galaxy itself and the possible polar ring look symmetric; 
their centers virtually coincide. These features
allow us to classify the object  as a "good candidate for a PRG" 
(by the terminology of Whitmore et al. 1990). Most of the previously 
studied nearby candidates for PRGs with similar morphology, e.g.
NGC\,660 (van Driel et al. 1995) and UGC 9562 (Reshetnikov and Combes 1994), 
actually turned out to be PRGs. This suggests that HUDF\,1619 is with high
probability a PRG. The main parameters of
HUDF\,1619 taken from the literature or found by us are collected in the table.

\begin{figure}[!ht]
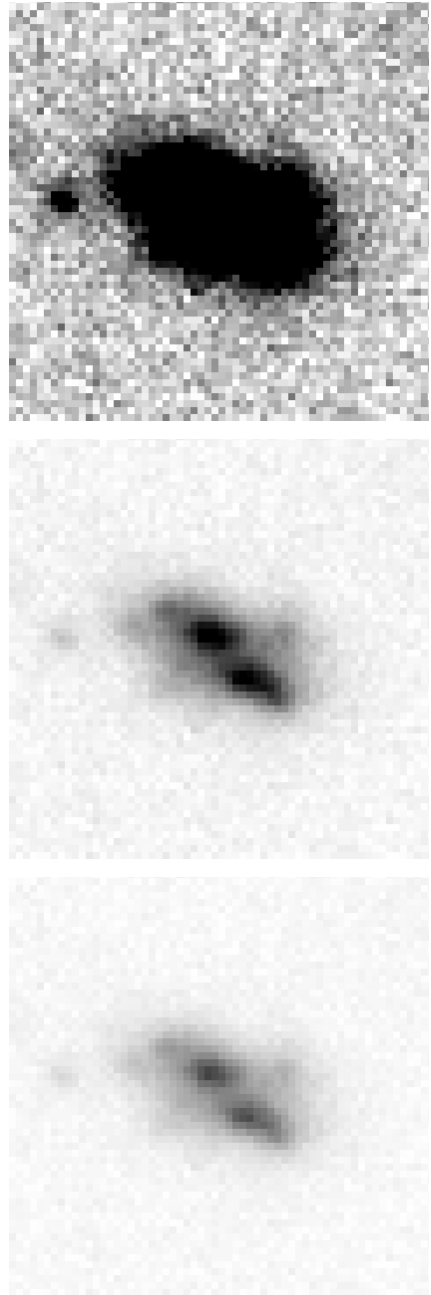

\centering{
\vspace*{-2.5cm}
\hspace*{7.0cm}
\vbox{
\includegraphics{fig1a.ps}
\includegraphics{fig1b.ps}
\includegraphics{fig1c.ps}
}\par
\vspace*{23.8cm}
\caption{HUDF\,1619 images with different contrasts to highlight its
main morphological features. The size of each frame is 
1.$''$8 $\times$ 1.$''$8. North is at the top and east is on the left.}
}
\end{figure}

\begin{figure}[!ht]
\centerline{\psfig{file=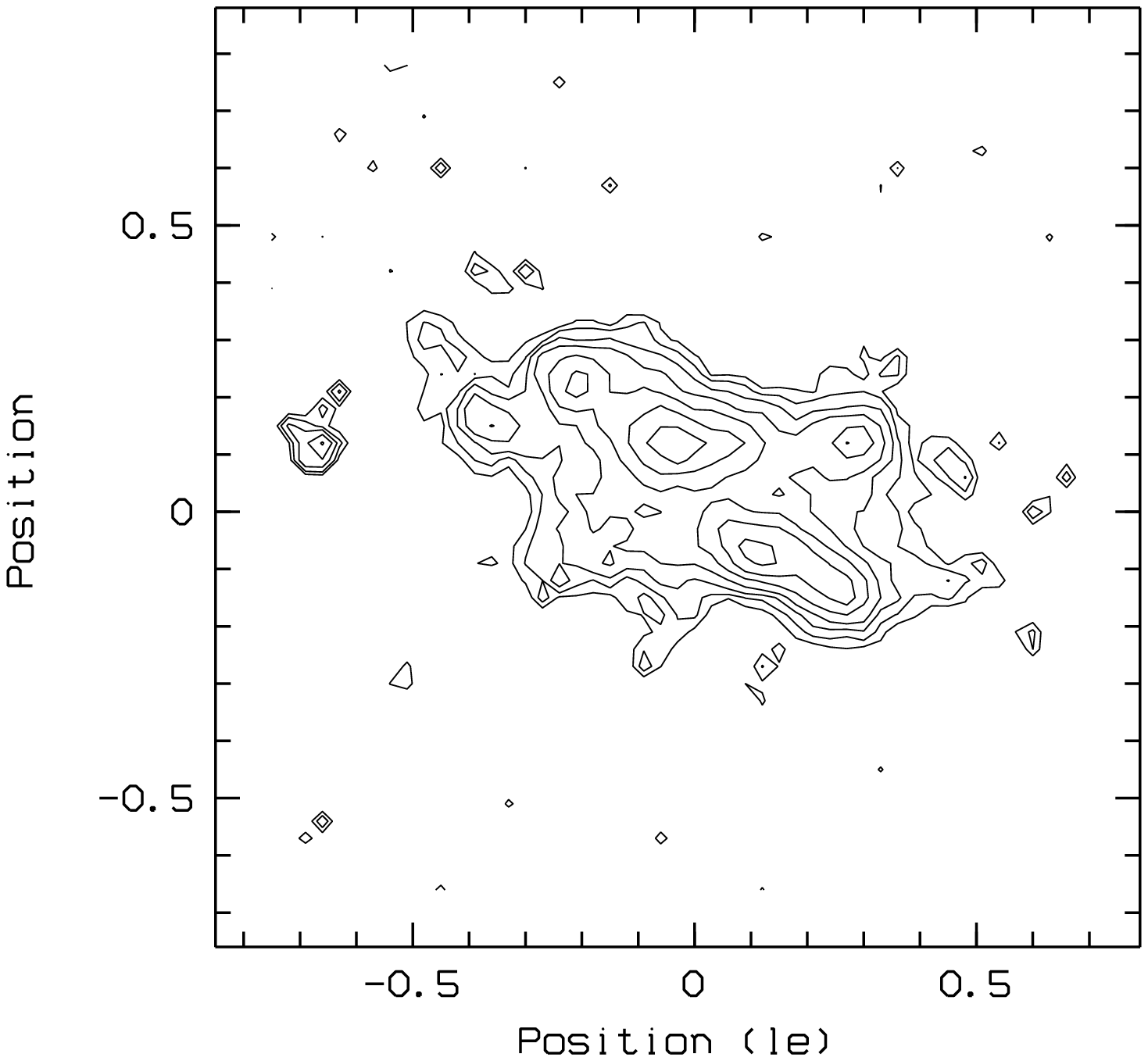,width=8.5cm,clip=,angle=-90}}
\caption{Isophotal map of HUDF\,1619 at 0.$^m$5 steps in the 
$i_{775}$ filter. The image of the galaxy was restored by Lucy's method 
(20 iterations). Arcseconds are along the axes.}
\end{figure}

\begin{table*}[!ht]
\caption{Main parameters of HUDF\,1619}
\begin{center}
\begin{tabular}{|c|c|c|}
\hline
Coordinates: $\alpha$(2000)      &  53.$^{\rm o}$1782428  & (1) \\
\hspace{2.5cm} $\delta$(2000)   & -27.$^{\rm o}$8076190  & (1) \\
Object's position in final HUDF frames:  &                &     \\
\hspace{3.6cm}  X (pixels)       & 3580.5                 & (1) \\
\hspace{3.6cm}  Y (pixels)       & 3305.6                 & (1) \\
Spectral type            &  Scd--Im   & (2)  \\
$B_{435}$                       & 25.38 &  (1) \\
$B_{435}-V_{606}$               & +0.19 &  (1) \\
$V_{606}-i_{775}$               & +0.32 &  (1) \\
$V_{606}-z_{850}$               & +0.87 &  (1) \\
Redshift ($z$)          &  1.3: &  (2) \\
Absolute magnitude ($M(B)$)& -20$^m$   &      \\
Optical diameter           &  1.$''$3 (11 kpc)  &      \\
Radial disk scalelength ($h_r$)      & 0.$''$13 (1.1 kpc) & \\
$L_{ring}/L_{galaxy}$ ($i_{775}$ filter) & 0.35 & \\
\hline
\end{tabular}
\end{center}
\begin{center}
(1) --  Beckwith et al. (2006) \\
(2) --  Coe et al. (2006)
\end{center} 
\end{table*}

    The galaxy has no published spectroscopic redshift. The project 
of slitless spectroscopy for HUDF objects (the Grism ACS Program for 
Extragalactic Surveys -- GRAPES; Pirzkal et al. 2004) provides
a spectrum of the galaxy, but it shows no distinct features that could 
be used to estimate $z$. The COMBO-17 catalog (Wolf et al. 2004) 
contains no estimate of the photometric $z$ for the galaxy, but it provides
the position of the peak in the probability distribution for possible values
of $z$. This peak lies at $z = 0.7$. On the
other hand, Coe et al. (2006) have recently given a
photometric redshift of 1.3 for HUDF\,1619. In what
follows, we assume the galaxy to be at $z = 1.3$. (If we
take $z = 0.7$, then all of the linear sizes mentioned
below should be reduced by 15\%\footnote{In this paper,
we take H$_0$=70 km/s/Mpc, $\Omega_m$=0.3, $\Omega_{\Lambda}$=0.7.}  
and the absolute magnitude of the galaxy 
with the $k$ correction applied will
increase, i.e., the galaxy will be fainter, by about 2$^m$).

    The outer disk of HUDF\,1619 demonstrates a
significant integral-shaped warp of the plane with
an amplitude of $\approx14^{\rm o}$ (see Figs. 1 and 2). Figure 3
shows photometric profiles of the galaxy along the
major axis of the central part of its main body\footnote{All magnitudes
here are given in the AB magnitude system.}. The
profiles exhibit a central dip in the brightness distribution that 
is associated, at least in part, with the
projection of the outer structure. At $\mid\,r\,\mid \geq 0.''2$, the 
observed brightness decreases approximately exponentially with the 
radial scalelength $h_r = 0.''13 \approx$ 1 kpc.
The color distribution over the disk of HUDF\,1619 is
asymmetric: its southwestern (SW) part is appreciably bluer.

   The observed vertical structure of the disk of the galaxy
can also be described by an exponential model with
$h_z\approx0.''04$. However, this is the observed value distorted 
by instrumental broadening. Since the FWHM sizes of the HUDF stellar 
images are 0.$''$08--0.$''$09 (Beckwith et al. 2006), the true vertical 
scale height should be smaller. (Recall that $h_z$ is not the size
of a particular structure, but a characteristic of the
slope of the brightness distribution.) Having modeled
the effect of instrumental broadening on the brightness distribution 
for HUDF galaxies of various sizes and thicknesses (this was done in 
the same way as in Reshetnikov et al. 2003), we found that $h_r$ for
HUDF\,1619 remains virtually unchanged, while the
vertical scale height should be reduced by a factor of 2.5, i.e. its 
undistorted value is $h_z\approx0.''015\approx0.13$ kpc. Consequently, 
the true ratio for the galaxy is $h_r/h_z = 9$, typical of nearby spiral 
galaxies (see, e.g., Fig. 6 in Reshetnikov et al. 2003).

    The observed luminosity of the possible inclined
ring structure reaches  $\sim$1/3 of the luminosity of the
central galaxy. The mean color indices of the ring are
close to those of the central galaxy, but their distribution is 
asymmetric: the southeastern (SE) part of the
ring is bluer than its northwestern part by 0.$^m$1--0.$^m$2
(this applies to all color indices).

\begin{figure}
\centerline{\psfig{file=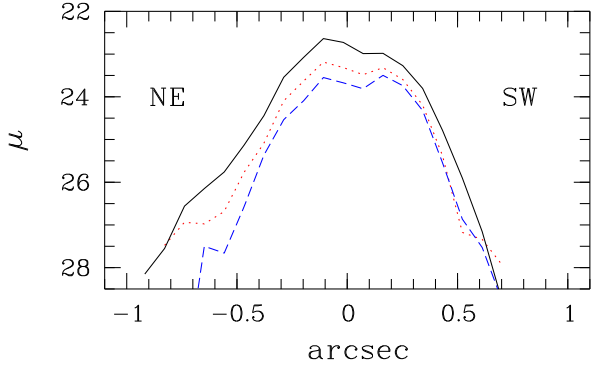,width=8cm,angle=-90,clip=}}
\caption{Photometric profiles of the galaxy along the major axis
of its central body (P.A.=40$^{\rm o}$). The solid, dotted , and
dashed curves are for the $z_{850}$, $i_{775}$, and $V_{606}$ filters,
respectively.}
\end{figure}

   Figure 4 shows the distribution of about two
thousand spiral and starburst galaxies in the color
($V_{606}$ -- $z_{850}$ and $B_{435}$ -- $V_{606}$) -- photometric 
redshift diagrams from Coe et al. (2006). The curves
indicate the color evolution models from Elmegreen and Elmegreen (2006). 
The models assume that the star formation began at $z = 6$ and the initial
metallicity was 0.008 (0.4$Z_{\odot}$). The star formation
lasted 1\,Gyr and the star formation rate decreased
over this time exponentially with a time scale of
0.3\,Gyr (upper curves) or remained constant (lower
curves). The color indices for most galaxies, including
HUDF\,1619 (large filled circles), are located between
the two curves, showing a trend in the color--$z$ plane
close to that expected in the model. Comparison of
the color indices for HUDF\,1619 with those of other
edge-on HUDF galaxies and with model calculations
shows that active star formation is under way and
young stars with ages $\leq$1 Gyr are present in the
galaxy itself and in the inclined ring (Elmegreen and
Elmegreen 2006).

\begin{figure}
\centerline{\psfig{file=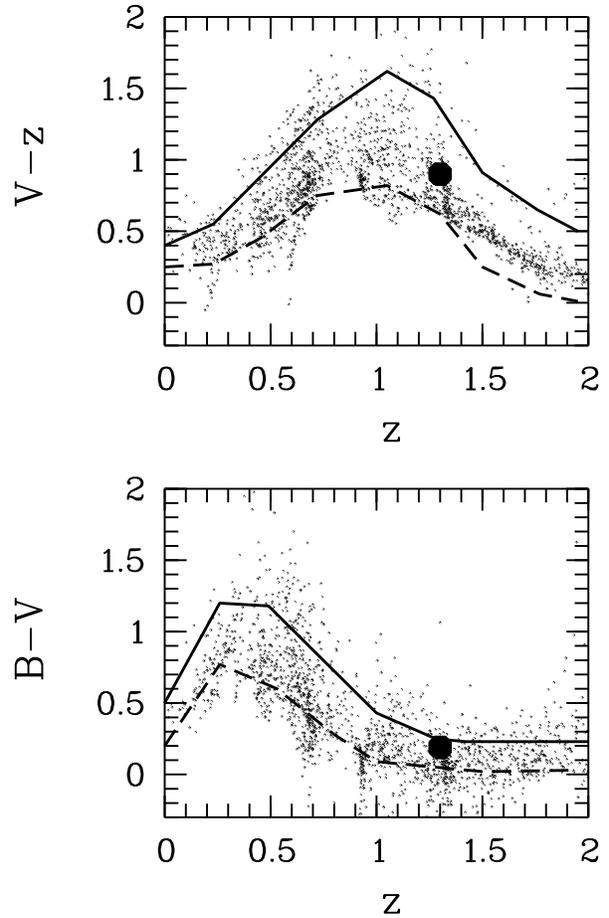,width=8cm,clip=,angle=-90}}
\caption{Color--photometric redshift diagram for spiral and
starburst galaxies from Coe et al. (2006) (dots). The filled
circles indicate the parameters of HUDF\,1619; the curves
represent the color evolution models.}
\end{figure}

\section{Discussion and conclusions}

   One of the most obvious PRG formation scenarios is the capture 
of matter in a circumpolar orbit during an encounter of two galaxies. 
Objects in the process of such a capture are observed directly
(e.g., NGC\,3808A,B) and, in addition, this scenario
has repeatedly been tested by numerical simulations
(Reshetnikov and Sotnikova 1997; Bournaud and Combes 2003; 
Reshetnikov et al. 2006). If the structure of HUDF\,1619 is explained 
by the interaction with another galaxy, then the donor galaxy should be
relatively close to HUDF\,1619.

    In a field with such a high projected density of
objects ($\sim$10\,000 galaxies or about one object per
$2'' \times 2''$ field are identified in the HUDF), candidates
for spatially close neighbors can be found near any
galaxy. Thus, for example, there is a strongly perturbed galaxy, 
HUDF\,1775, 4$''$ (or 33 kpc at $z = 1.3$) to the north of HUDF\,1619. 
Its apparent magnitude is comparable to that of HUDF\,1619 and it 
exhibits a structure resembling a highly bent tidal tail. However,
no spectroscopic $z$ value is available for this galaxy and,
just as for HUDF\,1619, the COMBO-17 catalog
(Wolf et al. 2004) gives no photometric $z$ estimate for
it. The $z$ estimate from Coe et al. (2006) ($z = 2.04$)
removes this galaxy from the list of potential donors,
however, this estimate may be inaccurate.

    A compact and slightly asymmetric galaxy
(HUDF\,1607), which is about 1$^m$ fainter than
HUDF\,1619 in the $V_{606}$ and $i_{775}$ filters, lies approximately 
2.$''$2 (18 kpc at $z = 1.3$) to the west of HUDF\,1619. According 
to Coe et al. (2006), $z = 1.3$ and, hence, this galaxy may be considered
as a candidate for a donor galaxy. To ultimately
clarify the question of whether HUDF\,1607 (and,
possibly, HUDF\,1775) is physically associated with
HUDF\,1619, accurate redshifts of the galaxies must
be determined.

    Thus, we detected a galaxy in the HUDF that
closely resembles PRGs observed in the local Universe. 
Previously, two more candidates for PRGs were identified in the 
Northern Hubble Deep Field (Reshetnikov 1997) at redshifts of
0.271 and 0.498 (Cohen et al. 2000). The Southern
Deep Field also contains objects that are morphologically similar to 
PRGs, but they can be attributed to this class with lesser confidence. 
Can any conclusions about the evolution of the PRG space density be 
reached if three objects are observed in three
deep fields? (Given the difficulty of classifying distant
PRGs, this number may be considered as a lower
limit for their actual number.)

    Let us use the parametrization of the space density
evolution in the form $(1 + z )^m$ and assume the local PRG density 
to be $3.5 \times 10^{-5}$ Mpc$^{-3}$ (this estimate was obtained by 
applying the standard $V/V_{max}$-method to the objects of categories 
A and B in the PRG catalog by Whitmore et al. 1990). Then, 
$m = 1.2$ will correspond to the three objects observed in
the $z$ range from 0.1 to 1.3 in the direction of the
three deep fields (their total area is $1.8 \times 10^{-6}$ sr). If,
however, the galaxy HUDF\,1619 is assumed to be
not at $z = 1.3$, but at $z = 0.7$ (see above), then our $m$
estimate will increase to 5.1. This may be indicative of
fast evolution of the PRG density.

   The above estimates show that the PRG statistics
can become a useful tool for estimating the rate of
interactions and mergers of galaxies at earlier epochs.
However, this requires both accurate, spectroscopic $z$
estimates and a further increase in the number of such
objects among distant galaxies.

\bigskip
\section*{Acknowledgments}
This work was supported by the Russian Foundation for
Basic Research (project no. 06-02-16459) and the German Academic
Exchange Service (DAAD).

\section*{REFERENCES}

\indent

 1. S.V.W.\,Beckwith, M.\,Stiavelli, A.M.\,Koekemoer,
    et al., Astron. J. 132, 1729 (2006).

 2. F.\,Bournaud and F.\,Combes, Astron. Astrophys. 401, 817 (2003).

 3. D.\,Coe, N.\,Benitez, S.F.\,Sanchez, et al., Astron. J. 132, 926 (2006).

 4. J.G.\,Cohen, D.W.\,Hogg, R.\,Blandford, et al., Astrophys. J. 538, 29 (2000).

 5. B.G.\,Elmegreen and D.M.\,Elmegreen, Astrophys. J. 650, 644 (2006).

 6. N.\,Pirzkal, C.\,Xu, S.\,Malhotra, et al., Astrophys. J., Suppl. Ser. 154, 
    501 (2004).

 7. V.P.\,Reshetnikov, Astron. Astrophys. 321, 749 (1997).

 8. V.P.\,Reshetnikov and F.\,Combes, Astron. Astrophys. 291, 57 (1994).

 9. V.\,Reshetnikov and N.\,Sotnikova, Astron. Astrophys. 325, 933 (1997).

10. V.P.\,Reshetnikov, R.-J.\,Dettmar, and F.\,Combes,
    Astron. Astrophys. 399, 879 (2003).

11. V.\,Reshetnikov, F.\,Bournaud, F.\,Combes, et al., Astron. Astrophys. 
    446, 447 (2006).

12. W.\,van Driel, F.\,Combes, F.\,Casoli, et al., Astron. J.
    109, 942 (1995).

13. B.C.\,Whitmore, R.A.\,Lucas, D.B.\,McElroy, et al.,
    Astron. J. 100, 1489 (1990).

14. C.\,Wolf, K.\,Meisenheimer, M.\,Kleinheinrich, et al.,
    Astron. Astrophys. 421, 913 (2004).\\

                               Translated by N. Samus'

\end{document}